\documentclass[10pt,twocolumn,letter]{IEEEtran}
\usepackage{times}
\usepackage{amsbsy}
\usepackage{latexsym}
\usepackage{amssymb}
\usepackage{amsmath}
\usepackage{color}
\usepackage[ansinew]{inputenc}
\usepackage[dvips ]{graphicx}
\input epsf
\usepackage{epic,eepic}
\usepackage{url}

\title {\huge Bidirectional MMSE Algorithms for Interference Mitigation in CDMA Systems over Fast Fading Channels
\hspace{-1 em}}

\author
{Patrick Clarke and Rodrigo C. de Lamare\\
Department of Electronics, University of York, York, UK, YO10 5DD\\
Email: pjc500@ohm.york.ac.uk, rcdl500@ohm.york.ac.uk \hspace{-1 em}}

\begin{document}

\maketitle

\begin{abstract}
This paper presents adaptive bidirectional minimum mean-square error
(MMSE) parameter estimation algorithms for fast-fading channels. The
time correlation between successive channel gains is exploited to
improve the estimation and tracking capabilities of adaptive
algorithms and provide robustness against time-varying channels.
Bidirectional normalized least mean-square (NLMS) and conjugate
gradient (CG) algorithms are devised along with adaptive mixing
parameters that adjust to the time-varying channel correlation
properties. An analysis of the proposed algorithms is provided along
with a discussion of their performance advantages. Simulations for
an application to interference suppression in DS-CDMA systems show
the advantages of the proposed algorithms.
\end{abstract}


\section{Introduction}
\label{Introduction}

Low-complexity reception and interference suppression is essential
in multiuser mobile systems if battery power is to be conserved,
data-rates improved, and quality of service enhanced. Conventional
adaptive schemes fulfil many of these requirements and have been a
significant focus of the research literature
\cite{Multiuser_detection_Verdu,adapt_filt_Haykin}. However, in
time-varying fading channels commonly associated with highly mobile
systems, adaptive techniques \textcolor{red}{encounter problems when
estimating and tracking the variations of parameters such as the
receive} filter and channel state information (CSI). The development
of cost-effective parameter estimation and tracking techniques for
highly dynamic channels remains a very challenging problem.

Existing strategies to enhance the performance of estimation
techniques include the use of optimized convergence parameters
\textcolor{red}{in} conventional adaptive algorithms to extend their
ability \textcolor{red}{to deal with fading and improve their
convergence and tracking performance}
\cite{tracking_improv_rls_Toplis,Varibale_ss_lms_gelfand,delamaresmf,clarke_smjio}.
However, the stability of adaptive step-sizes and forgetting factors
is a concern unless they are constrained to lie within a predefined
region \cite{Varibale_ss_lms_gelfand}. Furthermore, the fundamental
problem of demodulating the transmitted data symbols whilst
suppressing multiuser interference (MUI) remains. Approaches to
avoid and/or improve the tracking and estimation of fading
coefficients have been reported in
\cite{Honig,Chen,Diff_MMSE_Madhow,delamarespl07,jidf}. Although a
channel might be highly time-variant, adjacent fading coefficients
can be approximately equal and have a significant level of
correlation. These properties can be exploited to obtain a sequence
of faded symbols where the primary purpose of the receive filter is
to suppress interference and track the ratio between successive
fading coefficients, dispensing with the estimation of the fading
coefficients themselves. However, this scheme has a number of
limitations due to the fact that only one correlation time instant
is employed, which results in instability and a difficulty to track
highly time-variant signals.

In this paper, a bidirectional minimum mean-square error (MMSE)
based interference suppression scheme for highly-dynamic fading
channels is presented. The channel correlation between adjacent time
instants is exploited to improve the robustness, tracking and
convergence performances of existing adaptive schemes. Bidirectional
normalized least mean squares (NLMS) and conjugate gradient (CG)
algorithms are devised along with a mixing strategy that adaptively
weights the contribution of the considered time instants. An
analysis of the proposed schemes is given and establishes the
factors behind their behaviour and improved performance. The
proposed schemes are applied to DS-CDMA systems and simulations show
that they significantly outperform existing schemes.

The remainder of the paper is structured as follows, Section
\ref{sec:problem_statement} states the problem and explains the
motivation of the work. The algorithmic implementations of the
proposed bidirectional methods are given in Section
\ref{sec:proposed_algorithms}, and analysis of the proposed
algorithms in Section \ref{sec:algorithm_analysis}. Simulations and
performance evaluation are given in Section \ref{sec:Simulations}
and conclusions in Section \ref{sec:Conclusions}.

\section{DS-CDMA Signal Model and Problem Statement}
\label{sec:problem_statement}

Consider the uplink of a multiuser DS-CDMA system with $K$ users,
processing gain $N$ and multipath channels with $L$ paths. The
$M\times 1$ received signal after chip-pulsed matched filtering and
sampling at the chip rate is given by
\begin{equation}
\begin{split}
{\mathbf r}[i] & = A_{1}b_{1}[i]{\mathbf H}_{1}[i]{\mathbf c}_{1}[i]
+\underbrace{\sum^{K}_{k=2}A_{k}b_{k}[i]{\mathbf H}_{k}[i]{\mathbf
c}_{k}[i]}_{\rm{MUI}} \\ & \quad +{\boldsymbol \eta}[i] + {\mathbf
n}[i],
\end{split}
\end{equation}
where $M=N+L-1$, and ${\mathbf c}_{k}[i]$ and $A_{k}$ are the
spreading sequence and signal amplitude of the $k^{\mathrm{th}}$
user, respectively. The $M \times N$ matrix ${\mathbf H}_{k}[i]$
models the channel propagation effects for the $k^{\mathrm{th}}$
user, $b_k[i]$ corresponds to the transmitted symbol of the the
$k$th user, ${\boldsymbol \eta}[i]$ is the ISI vector and ${\mathbf
n}[i]$ is the noise vector. 

The design of linear receivers consists of processing the received
vector ${\boldsymbol r}[i]$ with the receive filter ${\boldsymbol
w}_{k}[i]$ with $M$ coefficients that provide an estimate of the
desired symbol as follows
\begin{equation}
z_k[i] = {\boldsymbol w}_{k}^{H}[i]{\boldsymbol r}[i],
\label{symbol}
\end{equation}
where the detected symbol is given by $\hat{b}_{k}[i] =
Q({\boldsymbol w}_{k}^{H}[i]{\boldsymbol r}[i])$, where ${\rm
Q}(\cdot)$ is a function that performs the detection according to
the constellation employed. It is also possible to use non-linear
receiver techniques. The problem we are interested in solving in
this work is how to estimate the parameter vector ${\boldsymbol
w}_{k}[i]$ of the receive filter in fast time-varying channels.

\section{Proposed Bidirectional Processing Strategy}

Adaptive parameter estimation techniques have two primary
objectives: estimation and tracking of the desired parameters.
However, in fast fading channels the combination of these two
objectives places unrealistic demands on conventional estimation
schemes. Differential techniques reduce these demands by relieving
the adaptive filter of the task of tracking fading coefficients.
This is achieved by posing an optimisation problem where the ratio
between two successive received samples is the quantity to be
tracked. Such an approach is enabled by the presumption that,
although the fading is fast, there is a significant level of
correlation between the adjacent channel samples
\begin{equation}
f_{1}[i]=E\left[h_{1}[i]h_{1}^{\ast}[i+1]\right]\geq 0,
\end{equation}
where $h_{1}[i]$ is the channel coefficient of the desired user. The
interference suppression of the resulting filter is improved in fast
fading environments compared to conventional adaptive filters but
only the ratio of adjacent fading samples is obtained. Consequently,
differential modulation, where the ratio between adjacent symbols is
the data carrying mechanism, are suited to differential MMSE
schemes.

However, limiting the optimization process to two adjacent samples
exposes the differential MMSE process to the negative effects of
uncorrelated samples
\begin{equation}
E\left[h_{1}[i]h_{1}^{\ast}[i+1]\right]\approx 0,
\end{equation}
and does not exploit the correlation that may be present between two or more adjacent samples, i.e.
\begin{equation}
\begin{array}{l}
f_{2}[i]=E\left[h_{1}[i]h_{1}^{\ast}[i-1]\right]>0 ~ {\rm and}~\\
f_{3}[i]=E\left[h_{1}[i+1]h_{1}^{\ast}[i-1]\right]>0.
\end{array}
\end{equation}
To address these weaknesses, we propose a bidirectional MSE cost
function based on adjacent received data vectors so that the number
of channel scenarios \textcolor{red}{under which an algorithm
performs reliable estimation and tracking is increased and
performance improved}. Termed the bidirectional MMSE, due to the use
of multiple and adjacent time instants, it can exploit the
correlation between successive received signals and reuse data
\cite{adapt_filt_Haykin}.

The optimization problem of the proposed scheme is given by
\begin{equation}
\begin{split}
{\boldsymbol w}_{\rm MMSE}  = \arg \min_{{\boldsymbol w}}  & ~E[
\sum^{D-2}_{d=0} \sum^{D-1}_{l=d+1} \rho_n[i]| b[i-d] {\boldsymbol
w}^H {\boldsymbol r}[i-l] \\ & \quad - b[i-l] {\boldsymbol w}^H
{\boldsymbol r}[i-d] |^2], \label{eq:Bidirectional_Cost_Function}
\end{split}
\end{equation}
where ${\boldsymbol w}_{\rm MMSE}$ is the expected value of the
filter, $\rho_n[i]$ is a weighting factor used in the cost function
to address problems with uncorrelated fading coefficients and
$n=d(D-3)+l+1$. Note that the time instants of interest have been
altered to avoid the use of future samples. In addition to
(\ref{eq:Bidirectional_Cost_Function}), an output power constraint
is required to avoid the trivial zero correlator solution
\begin{equation}
E\left[\left| {\boldsymbol w}^{H}{\boldsymbol
r}[i]\right|^{2}\right]=1. \label{eq:Non_Zero_Constraint}
\end{equation}
In fast-fading channels, the correlation between the considered time
instant is unlikely to be equal. Therefore variable weighting or
mixing of the cost function will be required to obtain improved
performance. However, the setting of the weights is problematic if
they are to be fixed. An adaptive scheme is preferable which can
take account of the time-varying channels. The errors extracted from
the cost function (\ref{eq:Bidirectional_Cost_Function}) are chosen
as the metric for this approach. This provides an input to the
weighting factor calculation process that is directly related to the
optimization in (\ref{eq:Bidirectional_Cost_Function}). The
time-varying mixing factors are given by
\begin{equation}
\rho_{n}[i] = \lambda_{e}\rho_{n}[i-1] + (1-\lambda_{e})\frac{e_{T}[i]-|e_{n}[i]|}{e_{T}[i]}
\label{eq:mixing_factors}
\end{equation}
where $ e_{T}[i] = |e_{1}[i]|+|e_{2}[i]|+ \ldots + |e_{D}[i]|$ and
the individual errors terms are calculated for $d=0,\ldots, D-2$ and
$l=d+1, \ldots, D-1$
\begin{equation}
e_n[i] =  b[i-d] {\boldsymbol w}^H[i] {\boldsymbol r}[i-l]- b[i-l]
{\boldsymbol w}^H[i] {\boldsymbol r}[i-d] . \label{eq:en}
\end{equation}
The forgetting factor, $0\leq \lambda_{e}\leq1$, is user defined
and, along with the normalization by the total error, $e_{T}[i]$,
and $\sum^{D}_{n=1}\rho_{n}[0]=1$,  ensures
$\sum^{D}_{n=1}\rho_{n}[i]=1$ and a convex combination at each time
instant.

\section{Proposed Bidirectional Algorithms}
\label{sec:proposed_algorithms}

In this section, the proposed bidirectional MMSE-based algorithms
based on (\ref{eq:Bidirectional_Cost_Function}) are derived. In
particular, we concentrate on the case where $D=3$ and develop
bidirectional NLMS and CG adaptive algorithms. Let us consider the
following cost function {
\begin{equation}
\begin{split}
\hspace{-1em} C({\boldsymbol w}[i],\rho_n[i]) & =
E\big[\rho_1[i]|b[i] {\boldsymbol w}^H[i] {\boldsymbol r}[i-1] \\ &
\quad -b[i-1]
{\boldsymbol w}^H[i] {\boldsymbol r}[i] |^2\\
& \quad +\rho_2[i]|b[i]{\boldsymbol w}^H[i]
{\boldsymbol r}[i-2] \\ & \quad -b[i-2]{\boldsymbol w}^H[i]{\boldsymbol r}[i] |^2\\
& \quad +\rho_3[i]|b[i-1]{\boldsymbol w}^H[i]
{\boldsymbol r}[i-2] \\ & \quad -b[i-2]{\boldsymbol w}^H[i]{\boldsymbol r}[i-1]|^2 \big]\\
\label{eq:cost3}
\end{split}
\end{equation}}
The time-varying mixing factors are adjusted by
\ref{eq:mixing_factors}
where $ e_{T}[i] = |e_{1}[i]|+|e_{2}[i]|+|e_{3}[i]|$ and the
individual errors terms are given by
\begin{equation}
\begin{split}
& e_{1}[i] = b[i]{\boldsymbol w}^{H}[i-1]{\boldsymbol r}[i-1]-b[i-1]{\boldsymbol w}^{H}[i-1]{\boldsymbol r}[i] \\
& e_{2}[i] = b[i]{\boldsymbol w}^{H}[i-1]{\boldsymbol r}[i-2]-b[i-2]{\boldsymbol w}^{H}[i-1]{\boldsymbol r}[i] \\
& e_{3}[i] = b[i-1]{\boldsymbol w}^{H}[i-1]{\boldsymbol
r}[i-2]-b[i-2]{\boldsymbol w}^{H}[i-1]{\boldsymbol r}[i-1].
\label{eq:mix3}
\end{split}
\end{equation}
The forgetting factor, $0\leq \lambda_{e}\leq1$, is user defined
and, along with the normalization by the total error, $e_{T}[i]$,
and $\sum^{3}_{n=1}\rho_{n}[0]=1$,  ensures
$\sum^{3}_{n=1}\rho_{n}[i]=1$ and a convex combination at each time
instant.

\subsection{Bidirectional NLMS Algorithm}

We first devise a low-complexity bidirectional NLMS algorithm that
iteratively computes the solution of (\ref{eq:cost3}). The
instantaneous gradient of (\ref{eq:cost3}) is taken with respect to
${\boldsymbol w}^{*}[i]$, and the errors terms of (\ref{eq:mix3})
are incorporated to yield the update equation
\begin{equation}
\begin{array}{ll}
{\boldsymbol w}[i] = &{\boldsymbol w}[i-1] + \frac{\mu}{M[i]}\left[\rho_{1}[i]b[i-1]{\boldsymbol r}[i]e_{1}[i] \cdots\right. \\
&\left.+ \rho_{2}[i]b[i-2]{\boldsymbol r}[i]e_{2}[i] +
\rho_{3}[i]b[i-2]{\boldsymbol r}[i-1]e_{3}[i]\right]
\end{array},
\label{eq:NLMS_Update_mixing}
\end{equation}
where $\mu$ is the step-size and the adaptive mixing parameters have
been included. The normalization factor, $M[i]$, is given by
\begin{equation}
M[i] = \lambda_{M}M[i-1] + (1-\lambda_{M}){\boldsymbol
r}^{H}[i]{\boldsymbol r}[i]
\end{equation}
where $\lambda_{M}$ is an exponential forgetting factor
\cite{Diff_MMSE_Madhow}. The enforcement of the constraint is
performed by the denominator of (\ref{eq:NLMS_Update_mixing}) and
ensures that the the filter ${\boldsymbol w}[i]$ does not tend
towards a zero correlator. The complexity of this algorithm is
$O(DM)$, which corresponds to roughly $D=3$ times that of the NLMS.

\subsection{Bidirectional Conjugate Gradient Algorithm}
Due to the incongruous form of the bidirectional formulation and the
conventional matrix inversion lemma based recursive least-squares
(RLS) algorithm, an alternative bidirectional CG algorithm is now
derived. We begin with the time-averaged autocorrelation and
crosscorrelation structures $\bar{\mathbf{R}}$ and
$\bar{\mathbf{t}}$ from (\ref{eq:cost3})
\begin{equation}
\begin{array}{l}
\bar{{\boldsymbol R}}_{1}[i] = \lambda\bar{{\boldsymbol R}}_{1}[i-1] + b[i-1]{\boldsymbol r}[i]{\boldsymbol r}^{H}[i]b^{\ast}[i-1]\\
\bar{{\boldsymbol R}}_{2}[i] = \lambda\bar{{\boldsymbol R}}_{2}[i-1] + b[i-2]{\boldsymbol r}[i]{\boldsymbol r}^{H}[i]b^{\ast}[i-2]\\
\bar{{\boldsymbol R}}_{3}[i] = \lambda\bar{{\boldsymbol R}}_{3}[i-1]
+ b[i-2]{\boldsymbol r}[i-1]{\boldsymbol r}^{H}[i-1]b^{\ast}[i-2]
\end{array}
\label{eq:Bidirect_LS_Autocorrelation}
\end{equation}
and
\begin{equation}
\hspace{-0.95em}\begin{array}{l}
\bar{{\boldsymbol t}}_{1}[i] = \lambda\bar{{\boldsymbol t}}_{3}[i-1] + b[i-1]{\boldsymbol r}[i]{\boldsymbol r}^{H}[i-1]{\boldsymbol w}[i-1]b^{\ast}[i]\\
\bar{{\boldsymbol t}}_{2}[i] = \lambda\bar{{\boldsymbol t}}_{2}[i-1] + b[i-2]{\boldsymbol r}[i]{\boldsymbol r}^{H}[i-2]{\boldsymbol w}[i-1]b^{\ast}[i]\\
\bar{{\boldsymbol t}}_{3}[i] = \lambda\bar{{\boldsymbol t}}_{3}[i-1]
+ b[i-2]{\boldsymbol r}[i-1]{\boldsymbol r}^{H}[i-2]{\boldsymbol
w}[i-1]b^{\ast}[i-1]
\end{array},
\label{eq:Bidirect_LS_Crosscorrelation}
\end{equation}
respectively. After some algebraic manipulations with the terms, the
final correlation structures are given by
\begin{equation}
\bar{{\boldsymbol R}}[i] = \rho_{1}[i]\bar{{\boldsymbol R}}_{1}[i] +
\rho_{2}[i]\bar{{\boldsymbol R}}_{2}[i] +
\rho_{3}[i]\bar{{\boldsymbol R}}_{3}[i] \label{eq:mixing_ls_auto}
\end{equation}
\begin{equation}
\bar{{\boldsymbol t}}[i] = \rho_{1}[i]\bar{{\boldsymbol t}}_{1}[i] +
\rho_{2}[i]\bar{{\boldsymbol t}}_{2}[i] +
\rho_{3}[i]\bar{{\boldsymbol t}}_{3}[i] \label{eq:mixing_ls_cross}
\end{equation}
where the adaptive mixing factors have been included. Inserting these structures into the standard CG quadratic form yields
\begin{equation}
J({\boldsymbol w}[i], \rho_n[i])= {\boldsymbol w}^{H}[i]{\boldsymbol
R}[i]{\boldsymbol w}[i]-{\boldsymbol t}^{H}[i]{\boldsymbol w}[i].
\label{eq:cg_quadratic_bidirect}
\end{equation}
From \cite{linear_non_linear_prog_luenburger}, the unique minimizer
of (\ref{eq:cg_quadratic_bidirect}) is also the minimizer of
\begin{equation}
{\boldsymbol R}[i]{\boldsymbol w}[i]={\boldsymbol t}[i].
\end{equation}
At each time instant, a number of iterations of the following method
are required to reach an accurate solution, where the iterations are
indexed with the variable $j$. At the $i^{\mathrm{th}}$ time instant
the gradient and direction vectors are initialized as
\begin{equation}
{\boldsymbol g}_{0}[i]=\nabla_{{\boldsymbol w}^*[i]}J({\boldsymbol
w}[i], \rho_n[i])={\boldsymbol R}[i]{\boldsymbol
w}_{0}[i]-{\boldsymbol t}[i]
\end{equation}
and
\begin{equation}
{\boldsymbol d}_{0}[i] = -{\boldsymbol g}_{0}[i],
\end{equation}
respectively, where the gradient expression is equivalent to those
used in the derivation of the NLMS algorithm. The vectors
${\boldsymbol d}_{j}[i]$ and ${\boldsymbol d}_{j+1}[i]$ are
${\boldsymbol R}[i]$ orthogonal with respect to ${\boldsymbol R}[i]$
such that ${\boldsymbol d}_{j}[i]{\boldsymbol R}[i]{\boldsymbol
d}_{l}[i]=0$ for $j\neq l$. At each iteration, the filter is updated
as
\begin{equation}
{\boldsymbol w}_{j+1}[i] = {\boldsymbol w}_{j}[i] +
\alpha_{j}[i]{\boldsymbol d}_{j}[i] \label{eq:CG_filter_update}
\end{equation}
where $\alpha_{j}[i]$ is the minimizer of $J_{LS}({\boldsymbol
w}_{j+1}[i])$ such that
\begin{equation}
\alpha_{j} = \frac{-{\boldsymbol d}_{j}^{H}{\boldsymbol
g}_{j}[i]}{{\boldsymbol d}_{j}^{H}[i]{\boldsymbol R}[i]{\boldsymbol
d}_{j}[i]}.
\end{equation}
The gradient vector is then updated according to
\begin{equation}
{\boldsymbol g}_{j+1}[i]={\boldsymbol R}[i]{\boldsymbol
w}_{j}[i]-{\boldsymbol t}[i]
\end{equation}
and a new CG direction vector found
\begin{equation}
{\boldsymbol d}_{j+1}[i] = -{\boldsymbol g}_{j+1}[i] +
\beta_{j}[i]{\boldsymbol d}_{j}[i]
\end{equation}
where
\begin{equation}
\beta_{j}[i] = \frac{{\boldsymbol g}_{j+1}^{H}[i]{\boldsymbol
R}[i]{\boldsymbol d}_{j}[i]}{{\boldsymbol d}_{j}^{H}[i]{\boldsymbol
R}[i]{\boldsymbol d}_{j}[i]} \label{eq:CG_beta_update}
\end{equation}
ensures the ${\boldsymbol R}[i]$ orthogonality between ${\boldsymbol
d}_{j}[i]$ and ${\boldsymbol d}_{l}[i]$ where $j\neq l$. The
iterations (\ref{eq:CG_filter_update}) - (\ref{eq:CG_beta_update})
are repeated until $j = j_{max}$.

\section{Analysis of the Proposed Algorithms}
\label{sec:algorithm_analysis}

The form of the bidirectional MSE cost function precludes the
application of standard MSE analysis. Consequently, we concentrate
on the signal to interference plus noise ratio (SINR) of the
proposed NLMS algorithm to analyze its performance.

\subsection{SINR Analysis}
To begin, we convert the SINR expression given by
\begin{equation}
\mathrm{SINR} = \frac{{\boldsymbol w}^{H}[i] {\boldsymbol
R}_{\mathrm{S}}{\boldsymbol w}[i]}{{\boldsymbol
w}^{H}[i]{\boldsymbol R}_{\mathrm{I}}{\boldsymbol w}[i]},
\end{equation}
where ${\boldsymbol R}_{S}$ and ${\boldsymbol R}_{I}$ are the signal
and interference and noise correlation matrices, into a form
amenable to analysis. Substituting in the filter error weight
vector, ${\boldsymbol \varepsilon }[i] = {\boldsymbol
w}[i]-{\boldsymbol w}_{o}[i]$, where ${\boldsymbol w}_{o}$ is the
instantaneous standard MMSE receiver, and taking the trace of the
expectation yields
\begin{equation}
\mathrm{SINR}=\frac{{\boldsymbol K}[i]{\boldsymbol R}_{\mathrm{S}}+
{\boldsymbol G}[i]{\boldsymbol R}_{\mathrm{S}}+
P_{\mathrm{S,opt}}[i]+ {\boldsymbol G}^{H}[i]{\boldsymbol
R}_{\mathrm{S}}} {{\boldsymbol K}[i]{\boldsymbol R}_{\mathrm{I}}+
{\boldsymbol G}[i]{\boldsymbol R}_{\mathrm{I}}+
P_{\mathrm{I,opt}}[i]+ {\boldsymbol G}^{H}[i]{\boldsymbol
R}_{\mathrm{I}}}, \label{eq:SINR_Analysis_2}
\end{equation}
where ${\boldsymbol K}[i] = E[{\boldsymbol
\varepsilon}[i]{\boldsymbol \varepsilon}^{H}[i]]$, ${\boldsymbol
G}[i]=E[{\boldsymbol w}_{o}[i]{\boldsymbol \varepsilon}^{H}[i]]$,
$P_{\mathrm{S,opt}}[i]=[{\boldsymbol w}_{o}^{H}[i]{\boldsymbol
R}_{\mathrm{S}}{\boldsymbol w}_{o}[i]]$ and
$P_{\mathrm{I,opt}}[i]=E[{\boldsymbol w}_{o}^{H}[i]{\boldsymbol
R}_{\mathrm{I}}{\boldsymbol w}_{o}[i]]$. From
(\ref{eq:SINR_Analysis_2}) it is clear that we need to pursue
expressions for ${\boldsymbol K}[i]$ and ${\boldsymbol G}[i]$ to
reach an analytical interpretation of the bidirectional scheme.

Substituting the filter error weight vector into the filter update
expression of (\ref{eq:NLMS_Update_mixing}) yields a recursive
expression for ${\boldsymbol \varepsilon}[i]$
\begin{equation}
\begin{array}{ll}
\hspace{-1.5em}{\boldsymbol \varepsilon}[i]& \hspace{-1 em}={\boldsymbol \varepsilon}[i-1]\\ &+\left[{\boldsymbol I}+\mu{\boldsymbol r}[i]b[i-1]{\boldsymbol r}^{H}[i-1]b^{\ast}[i]-\mu{\boldsymbol r}[i]b[i-1]{\boldsymbol r}^{H}[n]b^{\ast}[i-1]\right.\\ &+\mu{\boldsymbol r}[i]b[i-2]{\boldsymbol r}^{H}[i-2]b^{\ast}[i]-\mu{\boldsymbol r}[i]b[i-2]{\boldsymbol r}^{H}[n]b^{\ast}[i-2]\\ &+\mu{\boldsymbol r}[i-1]b[i-2]{\boldsymbol r}^{H}[i-2]b^{\ast}[i-1]\\
\hspace{-1.5em}&-\left.\mu{\boldsymbol r}[i-1]b[i-2]{\boldsymbol r}^{H}[i-1]
b^{\ast}[i-2]\right]{\boldsymbol \varepsilon}[i-1]\\ &
+\mu{\boldsymbol r}[i]b[i-1]e^{\ast}_{o,1}[i]
+\mu{\boldsymbol r}[i]b[i-2]e^{\ast}_{o,2}[i] \\ & +\mu{\boldsymbol r}[i-1]b[i-2]e^{\ast}_{o,3}[i]\\
\end{array},
\label{eq:bidirectional_stochastic_diff}
\end{equation}
where the terms $e_{o,1-3}$ are the error terms of (\ref{eq:mix3})
when the optimum filter ${\boldsymbol w}_{o}$ is used. Using the
direct averaging approach of Kushner
\cite{kushner_weak_convergence}, the solution to the stochastic
difference equation of (\ref{eq:bidirectional_stochastic_diff}) can
be approximated by the solution to a second equation
\cite{adapt_filt_Haykin}, such that
\begin{equation}
\begin{array}{l}
\hspace{-1.5em} E\left[{\boldsymbol I}+\mu{\boldsymbol r}[i]b[i-1]{\boldsymbol r}^{H}[i-1]b^{\ast}[i]-\mu{\boldsymbol r}[i]b[i-1]{\boldsymbol r}^{H}[n]b^{\ast}[i-1]\right.\\ +\mu{\boldsymbol r}[i]b[i-2]{\boldsymbol r}^{H}[i-2]b^{\ast}[i]-\mu{\boldsymbol r}[i]b[i-2]{\boldsymbol r}^{H}[n]b^{\ast}[i-2]\\ \left.+\mu{\boldsymbol r}[i-1]b[i-2]{\boldsymbol r}^{H}[i-2]b^{\ast}[i-1]\right.\\
-\left.\mu{\boldsymbol r}[i-1]b[i-2]{\boldsymbol r}^{H}[i-1]b^{\ast}[i-2]\right]\\
= {\boldsymbol I}+\mu{\boldsymbol F}_{1}-\mu{\boldsymbol
R}_{1}+\mu{\boldsymbol F}_{2}-\mu{\boldsymbol R}_{2}
+\mu{\boldsymbol F}_{3}-\mu{\boldsymbol R}_{3}
\end{array},
\label{eq:direct_ave_nlms_analysis}
\end{equation}
where ${\boldsymbol F}$ and ${\boldsymbol R}$ are correlations
matrices. Specifically, ${\boldsymbol R}_{1-3}$ are autocorrelation
matrices given by
\begin{equation}
\begin{array}{l}
{\boldsymbol R}_{1} = E\left[\mu{\boldsymbol r}[i]b^{\ast}[i-1]{\boldsymbol r}^{H}[i]b^{\ast}[i-1]\right]\\
{\boldsymbol R}_{2} = E\left[\mu{\boldsymbol r}[i]b^{\ast}[i-2]{\boldsymbol r}^{H}[i]b^{\ast}[i-2]\right]\\
{\boldsymbol R}_{3} = E\left[\mu{\boldsymbol
r}[i-1]b^{\ast}[i-2]{\boldsymbol r}^{H}[i-1]b^{\ast}[i-1]\right]
\end{array}
\end{equation}
and ${\boldsymbol F}_{1-3}$ cross-time-instant correlation matrices,
given by
\begin{equation}
\begin{array}{l}
{\boldsymbol F}_{1} = E\left[\mu{\boldsymbol r}[i]b^{\ast}[i-1]{\boldsymbol r}^{H}[i-1]b^{\ast}[i]\right]\\
{\boldsymbol F}_{2} = E\left[\mu{\boldsymbol r}[i]b^{\ast}[i-2]{\boldsymbol r}^{H}[i-2]b^{\ast}[i]\right]\\
{\boldsymbol F}_{3} = E\left[\mu{\boldsymbol
r}[i-1]b^{\ast}[i-2]{\boldsymbol r}^{H}[i-2]b^{\ast}[i-1]\right].
\end{array}
\end{equation}
Using (\ref{eq:direct_ave_nlms_analysis}) and the independence
assumptions of $E\left[e_{o,1-3}[i]{\boldsymbol
\varepsilon}[i]\right]=0$, $E\left[{\boldsymbol
r}^{H}[i]{\boldsymbol r}[i-1]\right]=0$ and
$E\left[b_{k}[i]b_{k}[i-1]\right]=0$, we arrive at an expression for
${\boldsymbol K}[i]$
\begin{equation}
\begin{array}{ll}
\hspace{-1.5em} {\boldsymbol K}[i]=&\left[{\boldsymbol I}+\mu{\boldsymbol F}_{1}-\mu{\boldsymbol R}_{1}+\mu{\boldsymbol F}_{2}-\mu{\boldsymbol R}_{2} +\mu{\boldsymbol F}_{3}-\mu{\boldsymbol R}_{3} \right]{\boldsymbol K}[i-1]\cdots\\
&\left[{\boldsymbol I}+\mu{\boldsymbol F}_{1}-\mu{\boldsymbol R}_{1}+\mu{\boldsymbol F}_{2}-\mu{\boldsymbol R}_{2} +\mu{\boldsymbol F}_{3}-\mu{\boldsymbol R}_{3} \right] \\
&+ \mu^{2}{\boldsymbol R}_{1}J_{min,1}[i] + \mu^{2}{\boldsymbol
R}_{2}J_{min,2}[i] + \mu^{2}{\boldsymbol R}_{1}J_{min,3}[i]
\end{array}
\end{equation}
where $J_{min,j}[i] = |e_{o,j}|^{2}$. Following a similar method, an
expression for ${\boldsymbol G}[i]$ can also be reached
\begin{equation}
{\boldsymbol G}[i] = {\boldsymbol G}[i-1]\left[\mu{\boldsymbol
F}_{1} - \mu{\boldsymbol R}_{1} + \mu{\boldsymbol F}_{2} -
\mu{\boldsymbol R}_{2} + \mu{\boldsymbol F}_{3} - \mu{\boldsymbol
R}_{3}\right].
\end{equation}
At this point we study the derived expression to gain an insight
into the operation of the bidirectional algorithm and the origins of
its advantages over the conventional differential scheme. Equivalent
expressions for the existing differential NLMS scheme are given by
\begin{equation}
\begin{array}{ll}
{\boldsymbol K}[i]=&\left[{\boldsymbol I}+\mu{\boldsymbol
F}_{1}-\mu{\boldsymbol R}_{1} \right]{\boldsymbol K}[i-1]
\left[{\boldsymbol I}+\mu{\boldsymbol F}_{1}-\mu{\boldsymbol R}_{1}\right] \\
&+ \mu^{2}{\boldsymbol R}_{1}J_{min,1}[i] \\
{\boldsymbol G}[i] = &{\boldsymbol G}[i-1]\left[\mu{\boldsymbol
F}_{1} - \mu{\boldsymbol R}_{1}\right].
\end{array}
\end{equation}
The bidirectional scheme has a number of additional correlation
terms compared to the existing scheme. Evaluating the
cross-time-instant matrices with regards to the independence
assumptions yields
\begin{equation}
\begin{array}{l}
\hspace{-1.0em} {\boldsymbol F}_{1} = |a_{1}|^{2}{\boldsymbol
c}_{1}{\boldsymbol
c}_{1}^{H}\underbrace{E\left[h[i]h^{\ast}[i-1]\right]}_{f_{1}[i]},~
{\boldsymbol F}_{2} = |a_{1}|^{2}{\boldsymbol c}_{1}{\boldsymbol
c}_{1}^{H}\underbrace{E\left[h[i]h^{\ast}[i-2]\right]}_{f_{2}[i]},\\~{\rm
and}~ {\boldsymbol F}_{3} = |a_{1}|^{2}{\boldsymbol
c}_{1}{\boldsymbol
c}_{1}^{H}\underbrace{E\left[h[i-1]h^{\ast}[i-2]\right]}_{f_{3}[i]}
\end{array}.
\label{eq:F_expressions}
\end{equation} \vspace{-0.25em}
From the expression above it is clear that underlying factor that
governs the SINR performance of the algorithms is the correlation
between the considered time instants, $f_{1-3}$ and similarity
between data-ruse and the use of $f_{1}$ and $f_{2}$. Accordingly,
it is the additional correlation factors of the bidirectional
algorithm that enhance its performance compared to the conventional
techniques, confirming the initial motivation behind the proposition
of the bidirectional approach. Lastly, the $f_{1-3}$ expressions of
(\ref{eq:F_expressions}) can be seen as the factors that influence
the number of considered time instants.

\begin{figure}[!htb]
\begin{center}
\def\epsfsize#1#2{1\columnwidth}
\epsfbox{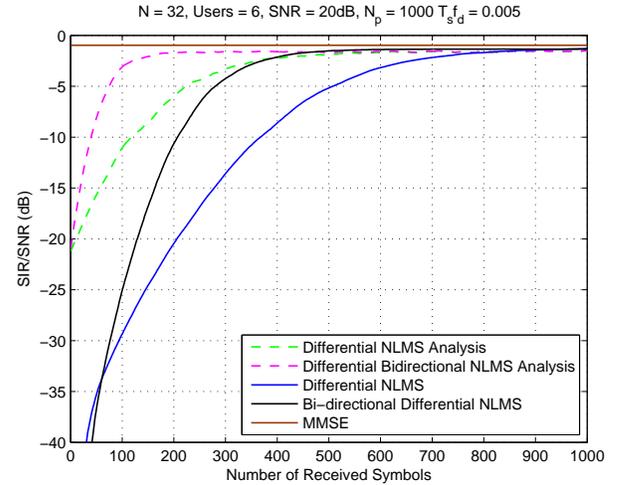} \vspace{-2em} \caption{\small Bidirectional
analytical SINR performance.}\label{fig:SINR}
\end{center}
\end{figure}

Central to the performance of the bidirectional schemes are the
correlation factors $f_{1-3}$ and the related assumption of
$h_{1}[i] \approx h_{1}[i-1]$. Examining the effect of the fading
rate on the value of $f_{1-3}$ shows that $f_{1}\approx f_{2}
\approx f_{3}$ at fading rates of up to $T_{s}f_{d}=0.01$, where
$T_{s}f_{d}$ is the normalized fading parameter. Consequently, after
a large number of received symbols with high total receive power
\begin{equation}
\hspace{-1.15em} 3\left[{\boldsymbol I}+\mu{\boldsymbol F}_{1} -
\mu{\boldsymbol R}_{1}\right]\approx\left[{\boldsymbol
I}+\mu{\boldsymbol F}_{1} - \mu{\boldsymbol R}_{1} + \mu{\boldsymbol
F}_{2} - \mu{\boldsymbol R}_{2} + \mu{\boldsymbol F}_{3} -
\mu{\boldsymbol R}_{3}\right], \label{eq:diff_bi_equivalence}
\end{equation}
due to the decreasing significance of the identity matrix.  This
indicates that the expected value of the SINR of the bidirectional
scheme, once $f_{1}\approx f_{2} \approx f_{3}$ have stabilized,
should be similar to the differential scheme. A second implication
is that the bidirectional scheme should converge towards the MMSE
level due to the equivalence between the bidirectional scheme and
the MMSE solution. Fig. \ref{fig:SINR} illustrates the analytical
performance using the above expressions.

The correlation matrices are calculated via ensemble averages prior
to commencement of the algorithm and ${\boldsymbol
G}[0]={\boldsymbol K}[0]={\boldsymbol I}$.  In Fig. \ref{fig:SINR}
one can see the convergence of the simulated schemes to the
analytical and MMSE plots, validating the analysis. Due to the
highly dynamic nature of the channel, using the expected values of
the correlation matrix alone cannot capture the true transient
performance of the algorithms. However, the convergence period of
the analytical plots within the first 200 iterations can be
considered to be within the coherence time and therefore give an
indication of the transient performance relative to other analytical
plots. Using this justification and the aforementioned analysis, it
is clear that the advantages brought by the bidirectional scheme are
predominantly in the transient phase due to the additional
correlation information supplied by ${\boldsymbol F}_{2}$ and
${\boldsymbol F}_{3}$ and their analogy with data reuse algorithms.
This observation is supported by the similar forms of the analytical
and simulated results and their subsequent convergence.

\section{Simulations}
\label{sec:Simulations}

We apply the proposed bidirectional adaptive algorithms to
interference suppression in the uplink of the DS-CDMA system
described in Section II. Simulations are averaged over $N_{p}$
packets and the parameters are specified in each plot. Conventional
schemes use BPSK modulation and the differential schemes employ
differential phase shift keying where the sequence of data symbols
to be transmitted by user $k$ are given by $b_{k}[i] =
a_{k}[i]b_{k}[i-1]$ where $a_{k}[i]$ is the unmodulated data.

\begin{figure}[!htb]
\begin{center}
\def\epsfsize#1#2{0.95\columnwidth}
\epsfbox{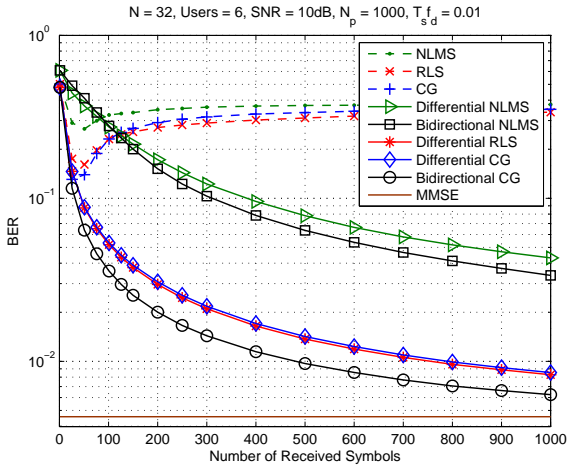} \vspace{-1em} \caption{\small BER performance
comparison of proposed schemes during training.}\label{fig:ber}
\end{center}
\end{figure}

The BER performance of existing and bidirectional schemes is
illustrated in Fig. \ref{fig:ber}. The existing RLS and proposed CG
algorithms converge to near the MMSE level with the bidirectional
scheme providing a clear performance advantage. However, the NLMS
schemes have a slower convergence performance due to their reduced
adaptation rate compared to the CG algorithms.

\begin{figure}[!htb]
\begin{center}
\def\epsfsize#1#2{0.95\columnwidth}
\epsfbox{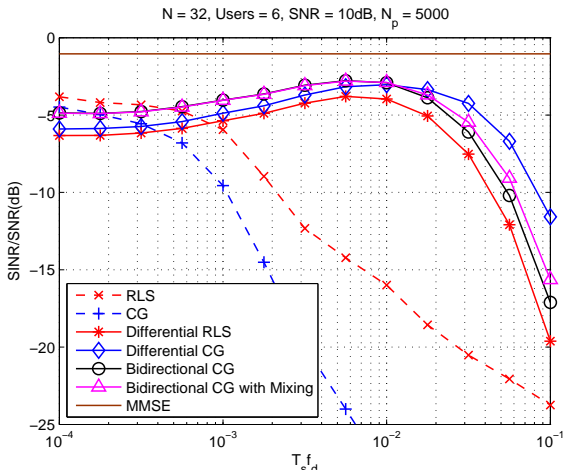} \vspace{-1em} \caption{\small SINR performance
versus fading rate of the proposed CG schemes after 200 training
symbols.}\label{fig:SINR2}
\end{center}
\end{figure}

Fig. \ref{fig:SINR2} illustrates the performance of the proposed CG
and existing RLS algorithms as the fading rate is increased, where
the SINR is normalised by the instantaneous SNR. The conventional
schemes are unable to cope with fading rates in excess of
$T_{s}f_{d}=0.005$  and begin to diverge at the completion of the
training sequence. The bidirectional scheme outperforms the
differential schemes but the performance begins to decline once
fading rates above $f_{d}T_{s}=0.01$ are reached. The increase in
performance of the bidirectional scheme can be accounted for by the
increased correlation information supplied by the matrices
${\boldsymbol F}_2$ and ${\boldsymbol F}_{3}$ and effective data
reuse. A second benefit of the bidirectional scheme is the improved
performance at low fading rate. The introduction of the mixing
factors into the bidirectional algorithm improves performance
further, especially at higher fading rates. An explanation for this
can be established by referring back to the observations on the
correlation factors $f_{1-3}$. Although fading rates of $0.01$ may
be fast, the assumption $h[i-2]\approx h[i-1]\approx h[i]$ is still
valid. Consequently, $f_{1}\approx f_{2} \approx f_{3}$ and equal
weighting is optimum. However, as the fading rate increases beyond
$T_{s}f_{d}=0.01$ this assumption breaks down and the correlation
information requires unequal weighting for optimum performance, a
task fulfilled by the adaptive mixing factors.

\section{Conclusions}
\label{sec:Conclusions}

We have presented bidirectional MMSE-based parameter estimation
algorithms that exploit the time correlation of rapidly varying
fading channels. The ratio between successive received vectors is
tracked using correlation information gathered at adjacent time
instants to avoid tracking of the faded or unfaded symbols. The
results show that the proposed algorithms applied to interference
suppression in DS-CDMA systems significantly outperform existing
algorithms.


\end{document}